\documentclass[11pt]{article}
\usepackage{amssymb}
\usepackage{graphicx}
\newcommand{\slabel}[1]{\label{#1}}
\newcommand{\psx}{\mathbf{x}}
\newcommand{\qsx}{\mathbf{X}}
\newcommand{\isu}{\hat\mathbf{u}}
\newcommand{\hsu}{\hat\mathbf{h}}
\newcommand{\qspace}{\mathbb{Q}}
\newcommand{\pspace}{\mathbb{P}}
\newcommand{\sspace}{\mathbb{S}}
\newcommand{\dspace}{\mathbb{D}}
\newcommand{\ispace}{\mathbb{I}}
\newcommand{\pp}{\hfill\par \vspace{\baselineskip}}
\newcommand{\pnorm}[2]{{|\!| #1 |\!|}_{\, #2}}
\newcommand{\bcdot}{\mathbf{\,\cdot\,}}

\begin{document}

\begin{center}
\textbf{MULTIPARTICLE CORRELATIONS IN Q-SPACE}\footnote{in: 
\textit{Correlations and Fluctuations in QCD},
    Proceedings of the 10th International Workshop on Multiparticle
    Production, Crete, June 2002, edited by N.G.\ Antoniou, F.K.\ 
    Diakonos and C.N.\ Ktorides,
    World Scientific (2003), pp. 386--403.}\\[8mm]

Hans C.\ Eggers\\
  \textit{Department of Physics, University of Stellenbosch}\\
  \textit{7600 Stellenbosch, South Africa}\\
  \ \\
  Thomas A.\ Trainor\\
  \textit{CENPA 354290, University of Washington} \\
  \textit{Seattle, WA 98195, USA}\\

\begin{quote}
{\small \textbf{Abstract:}
  We introduce $Q$-space, the tensor product of an index
  space with a primary space, to achieve a more general mathematical
  description of correlations in terms of $q$-tuples.  Topics
  discussed include the decomposition of $Q$-space into a sum-variable
  (location) subspace $\sspace$ plus an orthogonal difference-variable
  subspace $\dspace$, and a systematisation of $q$-tuple size
  estimation in terms of $p$-norms. The ``GHP sum'' prescription for
  $q$-tuple size emerges naturally as the 2-norm of difference-space
  vectors. Maximum- and minimum-size prescriptions are found to be
  special cases of a continuum of $p$-sizes.}
\end{quote}

\end{center}

\section{Correlations in P-space}
\label{sec:pspace}

Traditionally, particles emitted by high-energy hadronic or heavy ion
collisions have been visualised in terms of a collection of points
populating what we call \textit{primary space} $\pspace$ or P-space,
with each particle $i$ represented by a $d$-dimensional vector $\psx_i
= (x_{i1},x_{i2},\ldots,x_{id})$.  Examples of P-spaces are
three-momentum, with $\psx_i = (p_{ix},p_{iy},p_{iz})$ and
rapidity-azimuth, with $\psx_i = (y_i,\phi_i)$.  Particle correlations
can be studied either by binning $\pspace$ with a suitable partition
(``coarse-graining''), or by analysing distributions of relative
distances between primary vectors $\psx_i$ directly.  In this
contribution, we focus exclusively on the latter approach.
\pp

\textit{Correlations} are a matter of definition; typically they are
deviations of joint $q$-particle distributions from a pre-defined null
hypothesis or \textit{reference process}\cite{Egg01a} such as a
uniform distribution or a $q$-fold convolution of the differential
one-particle distribution. Under a ``dilute-fluid'' assumption that
correlation strength decreases with particle cluster ($q$-tuple) size,
low-order $q$-tuples --- particle pairs, triplets, quartets etc. --- are
usually studied as a first approximation, with higher-order $q$-tuples
as perturbations. In general, one selects out of $N$ particles all
possible combinations\footnote{
  In analytical manipulations, it is more convenient to consider all
  $N!/(N-q)!$ \textit{ordered} $q$-tuples rather than the $N!/(N-q)!
  q!$ \textit{unordered} ones.}
of $q$-tuples ($q = 1,2,3,\ldots$) for statistical analysis, using for
example \textit{cumulants}\cite{Stu87a} as differential correlation
measures. The characterisation of $q$-tuples is therefore fundamental
to correlation analysis.
\pp

The simplest properties of a given $q$-tuple are its location and
size. \textit{Location} can be defined, for example, as the centre of
mass of the $q$ particles,
\begin{equation}
\slabel{psb}
\overline{\psx} = \frac{1}{q} \sum_{i=1}^q \psx_i \,,
\end{equation}
or as the location of any one of the particles. 
\pp

\begin{figure}[htb]
  \centerline{
    \includegraphics[scale=0.27]{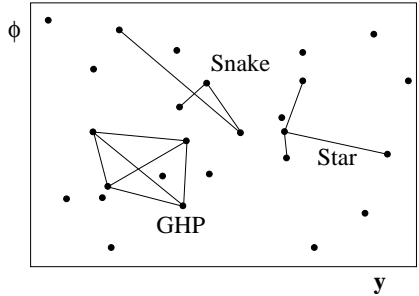} 
  }
\caption{An event, shown as a collection of $N$ points in the P-space 
  $(y,\phi)$. Also shown are examples of three topologies used to
  quantify $4$-tuple size.}
\label{fig:pspace}
\end{figure}

In its simplest incarnation, the mathematical realisation of
\textit{size} should be a nonnegative real number which reduces to
zero whenever all $q$ particles occupy the same position in $\pspace$:
in other words, size should be a \textit{norm based on relative
  coordinates.} Contrary to naive expectation, the best prescription
for this is not immediately obvious. While a 2-tuple's size is clearly
described in terms of the distance $|\psx_i - \psx_j|$, $q$-tuples of
higher order permit a range of choices. In Ref.\cite{Lip92a} this
problem was addressed in terms of different ``topologies'', summarised
pictorially in Fig.~\ref{fig:pspace} for some representative 4-tuples.
\pp

For every topology, interpair distances can be combined in several
ways to form different size estimators. The $q(q-1)/2$ interpair
distances making up the GHP topology can, for example, be summed to
yield the \textit{GHP sum} size; or size can be based on the largest
of these distances, resulting in the \textit{GHP max} size estimate.
Particle number within $d$-dimensional spheres centered on individual
particles defines the \textit{Star max} size, while the \textit{Snake
  Integral} seeks to quantify size in terms of a linear succession of
distances between ordered points.
\pp

Historically, the GHP max and Star max prescriptions were heuristic
inventions within the \textit{correlation integral} literature:
the $q{=}2$ simplest case\cite{Gra83a} was extended in 
Refs.\cite{Hen83a} 
and\cite{Gra83b} 
to the GHP max prescription, 
\begin{equation}
\slabel{psc}
C_q^{{\rm (GHP)}} (\ell) \equiv
\frac{1}{N^q} 
\, \{ \mbox{No.\ q-tuples\ } 
(i_1,\cdots,i_q)
\mbox{\ with all\ }
  |\psx_{i_n} - \psx_{i_m}| < \ell \}
\,,
\end{equation}
and in Refs.\cite{Pal84a,Paw87a,Atm89a} to the Star max,
\begin{equation}
\slabel{psd}
C_q^{{\rm (StarM)}} (\ell) \equiv
 \frac{1}{N^q} \sum_{i=1}^N 
     \left( \sum_{j=1}^N \theta(\ell - |\psx_i - \psx_j|\,) 
     \right)^{q-1}
\,,
\end{equation}
(with $\theta(x)$ the Heaviside function), which can, on multiplying
out the $(q{-}1)$ inner sums and using $\prod_j \theta(\ell - |\psx_i
- \psx_j|\,) = \theta(\ell - \max_j(|\psx_i - \psx_j|\,))$, be made to
exhibit the Star max prescription on the $q$-tuple
$(i,j_1,\cdots,j_{q-1})$ explicitly,
\begin{equation}
\slabel{pse}
C_q^{{\rm (StarM)}} (\ell) =
 \frac{1}{N^q}
    \sum_{i=1}^N \;\; \sum_{j_1,\cdots,j_{q-1}=1}^N 
     \theta(\ell - \max_{k}\{\, |\psx_i - \psx_{j_k}|\,\}\,) 
\,.
\end{equation}
Factorial extensions were published in Ref.\cite{Lip92a}.  Note that,
in summing indiscriminately over all $q$-tuples, the above correlation
integrals implicitly assume that correlation structure is independent
of location.

\section{Q-space by example}

\begin{figure}[htb]
  \centerline{
    \includegraphics[scale=0.32]{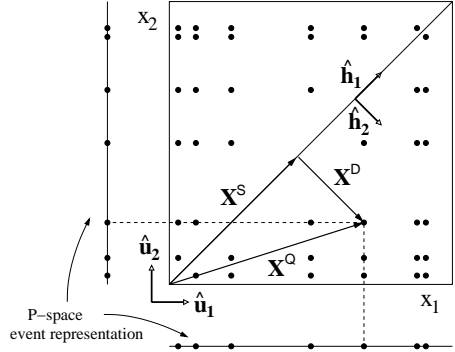} 
    }
\caption{Example of Q-space for $q{=}2, d{=}1$. The primary space
  event, represented on the margins, maps onto particle pairs in the
  Q-space square. Vector $\qsx^Q$, representing a particular pair,
  decomposes into a location vector $\qsx^S$ and a size vector
  $\qsx^D$. Unit vectors $\isu_{1,2}$ are rotated to $\hsu_1$ and
  $\hsu_2$ which span sum space and difference space respectively.}
\label{fig:rho2b}
\end{figure}

The concept of Q-spaces is not new: In high-energy hadronic
collisions, two-particle correlations have long been visualised in
two-particle spaces\cite{Foa75a}. Figure~\ref{fig:rho2b} illustrates
by means of a simple example how a Q-space is constructed from
P-space. A typical event in a one-dimensional ($d{=}1$) primary space
is represented by the dots on the lines below the $\psx_1$ axis and to
the left of the $\psx_2$ axis. Particle pairs are then represented by
all possible dots in the $Q{=}2$ space as shown by representative
dashed lines.\footnote{
  Making up a ``pair'' from a particle with itself would result in a
  dot lying on the diagonal line. Such usage corresponds to the
  transition from factorial to ordinary statistics. We ignore
  associated issues in this contribution.}
\pp

Each pair in this example is represented by a vector $\qsx^Q$ as
shown. The location corresponds to the component vector $\qsx^S$ along
the diagonal, and the 2-tuple size to the magnitude of $\qsx^D$, since
$|\qsx^S| = |x_1 + x_2|/\sqrt{2} = \sqrt{2}\,\overline{\psx}$ and
$|\qsx^D| = |x_1 - x_2|/\sqrt{2}$.  This can be derived algebraically
by using the Q-space representation $\qsx^Q = x_1 \isu_1 + x_2 \isu_2$
in terms of the \textit{index unit vectors} $\isu_{1,2}$, which are
rotated to the basis vectors $\hsu_{1,2}$ shown in
Fig.~\ref{fig:rho2b} to find $\qsx^S = \hsu_1\,( \hsu_1 \bcdot
\qsx^Q)$ and
$\qsx^D = \hsu_2 \, (\hsu_2 \bcdot \qsx^Q)$.
\pp

\begin{figure}
  \hfil
  \parbox[c]{55mm}{
    \includegraphics[scale=0.30]{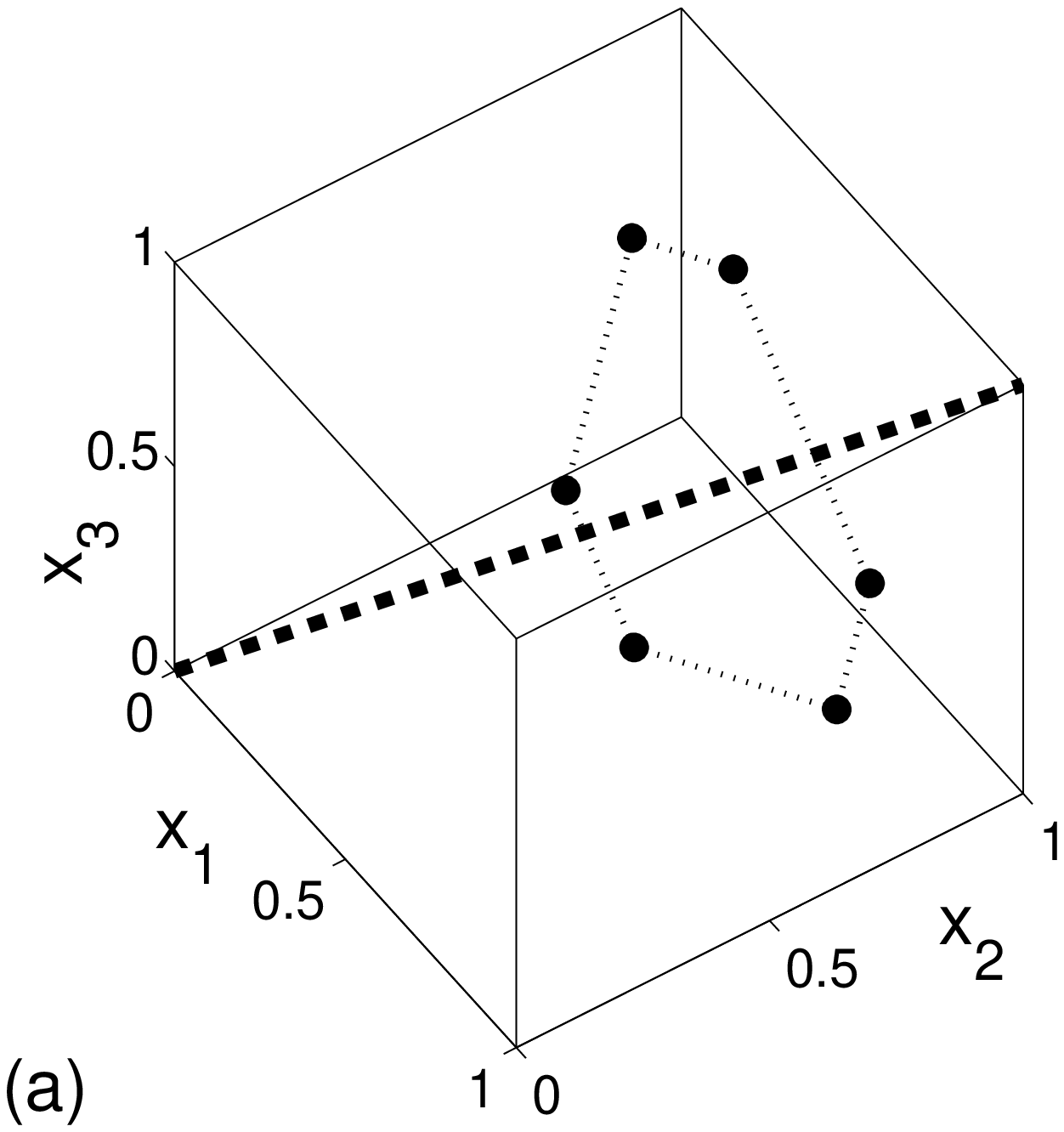}
    }
  \hfil
  \parbox[c]{45mm}{
    \includegraphics[scale=0.30]{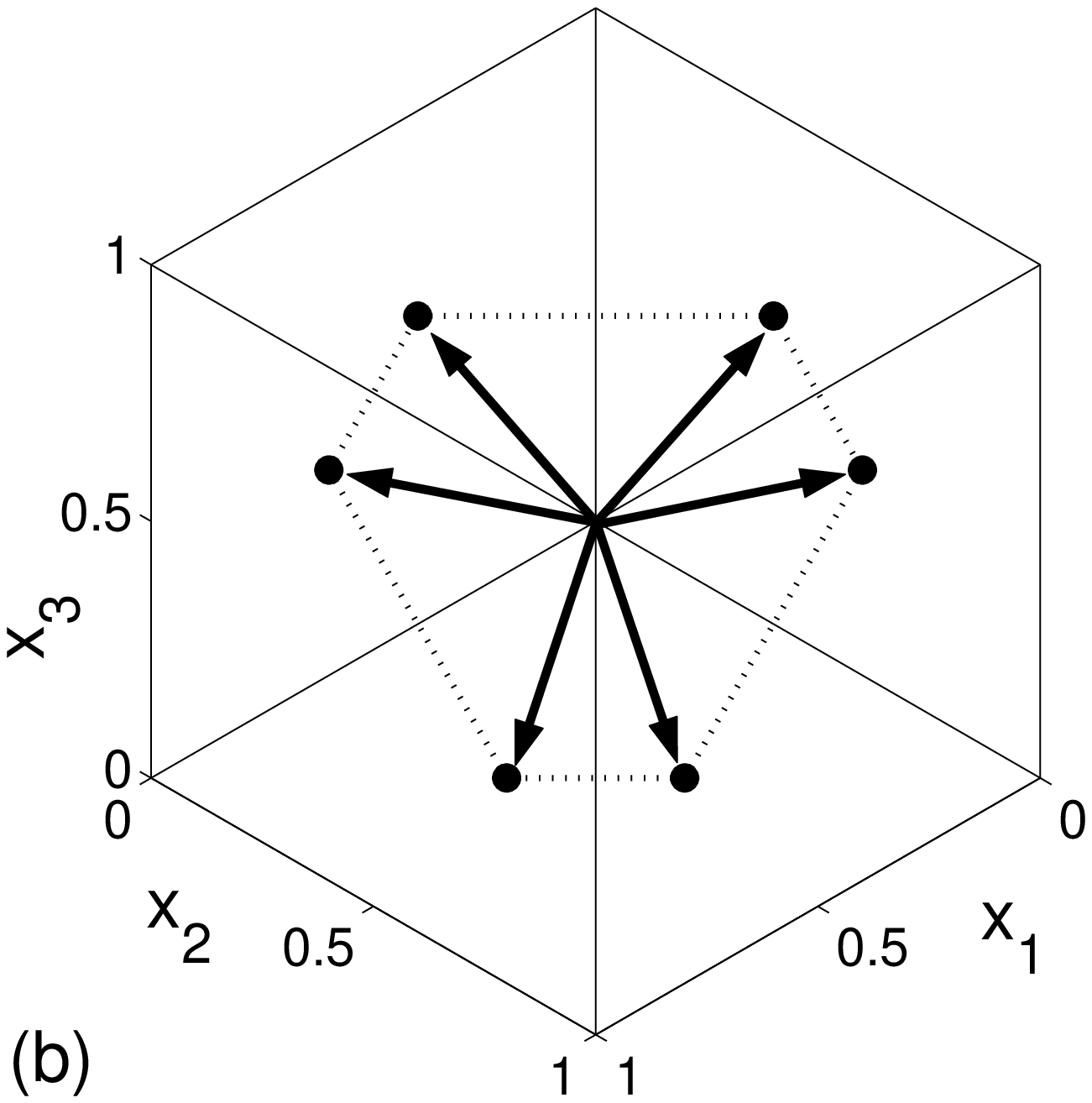}
    }
  \hfil\\
  \caption{A $3$-tuple (with $d{=}1$) showing (a) a $3$-tuple
    point plus its five permutation-symmetric counterparts and the
    main diagonal (dashed line), and (b) the same system when viewed
    down the main diagonal.  Also shown are the corresponding
    difference vectors $\qsx^D$. The dotted lines are coplanar with
    the \textit{difference space} $\dspace_3$.}
  \label{fig:threetuples}
\end{figure}

Correspondingly, 3-tuples can be visualised for one-dimensional
P-spaces as shown in Fig.~\ref{fig:threetuples}. Here, one particular
ordered triplet is shown in all its exchange-permutation incarnations,
all of which represent the same unordered triplet; the associated
symmetry is clearly visible when viewing the Q-space down the main
diagonal as in Fig.~\ref{fig:threetuples}(b).  The set of unit vectors
$(\isu_1,\isu_2,\isu_3)$ is transformed to
\begin{eqnarray}
\label{threeh}
\hsu_1 &=& (\isu_1 + \isu_2 + \isu_3)/\sqrt{3} \\
\hsu_2 &=& (\isu_1 - \isu_2)/\sqrt{2}\\
\hsu_3 &=& (\isu_1 + \isu_2 - 2 \isu_3)/\sqrt{6}
\end{eqnarray}
where $\hsu_1$ points along the main diagonal (shown as the dashed
line), while $\hsu_2$ and $\hsu_3$ span the plane, indicated by the
dotted lines in Fig.~\ref{fig:threetuples}, which is normal to the
main diagonal.  This normal plane exemplifies \textit{difference
  space} $\dspace_3$, within which only relative coordinates appear,
while the main diagonal represents the \textit{sum space} $\sspace_3$
measuring, once again, $q$-tuple location.
\pp

\section{Formalism for Q-space}

The general formalism for vectors in $\qspace$ is now easily
understood.  The particle vectors $\psx_i, i = 1,\ldots q$ of a
$q$-tuple in $d$-dimensional primary space $\pspace$ are combined into
the Q-space-vector
\begin{equation}
\qsx^Q = \sum_{i=1}^q \psx_i\,\isu_i \,.
\end{equation}
$\qspace$ is spanned by unit vectors which are the product of unit
vectors $\isu_i, i = 1,\ldots, q$ living in \textit{index-space}
$\ispace_q$ and the basis vectors of $\pspace$ (implicit in $\psx_i$).
The $d$-dimensional \textit{sum space} $\sspace_q$ is spanned by
\begin{equation}
\hsu_1 = \frac{1}{\sqrt{q}}\;\sum_{i=1}^q \isu_i
\end{equation}
and the basis vectors of $\pspace$. The corresponding index-space
projection of $\qsx^Q$ onto sum space, $\qsx^S \equiv \hsu_1\, (\hsu_1
\bcdot \qsx^Q)$, is given by
\begin{equation}\slabel{bsq}
\qsx^S
= \frac{1}{\sqrt{q}}\sum_{i=1}^q \psx_i
= \sqrt{q} \; \overline{\psx} \,,
\end{equation}
where $\overline{\psx}$ is the $q$-tuple CMS as before. Furthermore,
the algebraic complement of $\sspace_q$ is spanned by the set of
$(q-1)$ orthonormal vectors\footnote{
  Any basis set connected to this one by an orthogonal transformation
  will also do.}
in $\ispace_q$
\begin{eqnarray}
\slabel{bsra}
\hsu_2 &=&  (\isu_1 - \isu_2)/\sqrt{2}\,,\\
\slabel{bsrb}
\hsu_3 &=& (\isu_1 + \isu_2 - 2\isu_3)/\sqrt{6}\\[-1mm]
\vdots && \nonumber \\[-2mm]
\slabel{bsrc}
\hsu_q &=& \frac{1}{\sqrt{q(q-1)}}\; 
\sum_{i=1}^{q-1} (\isu_i - \isu_q) \,,
\end{eqnarray}
which, together with the $\pspace$ basis, are used to define a basis
for \textit{difference space} $\dspace_q$. The index space projection
of $\qsx^Q$ onto $\dspace_q$ is given by
\begin{equation}\slabel{bst}
\qsx^D 
\equiv \sum_{i=2}^q  \hsu_i \left(\hsu_i \bcdot \qsx^Q\right) \,.
\end{equation}
Since every $\qsx^Q \in \qspace$ has a unique decomposition into
$\qsx^S$ and $\qsx^D$, $\qspace$ is the direct sum of $\sspace_q$ and
$\dspace_q$. The relationship between the different spaces can hence
be summarised as
\begin{equation}\slabel{bsu} 
\ispace_q \otimes \pspace = \qspace
= \sspace_q \oplus \dspace_q \,,
\end{equation}
corresponding to the dimensionality relation $\dim(\ispace_q)\times
\dim(\pspace) = \dim(\qspace) = \dim(\sspace_q) + \dim(\dspace_q)$ \ 
or, explicitly,\ \ $q\times d = qd = d + (q-1)d$.
\pp

Within this framework, the first and most obvious measure of size is
the 2-norm of $\qsx^D$. Starting either with Eq.~(\ref{bst}) or by
subtraction, $\qsx^D = \qsx^Q - \qsx^S$, we find the suggestive form
\begin{equation}
  \slabel{szc}
\qsx^D  
= \sum_{i=1}^q \left( \psx_i - \overline{\psx} \right) \isu_i,
\end{equation}
which yields
\begin{equation}
  \slabel{szd}
  \left| \qsx^D \right|
= \frac{1}{\sqrt{q}}
  \left[ \; \sum_{i < j =1}^q \left( \psx_i - \psx_j\right)^2
  \right]^{1/2} \,,
\end{equation}
which, apart from the prefactor, is exactly the \textit{GHP 2-sum} or
\textit{rms} size measure. The GHP 2-sum thus appears to be the
natural measure of size in Q-space.

\section{Generalised $p$-sizes}

\begin{figure}[htb]
  \centerline{
      \includegraphics[scale=0.55]{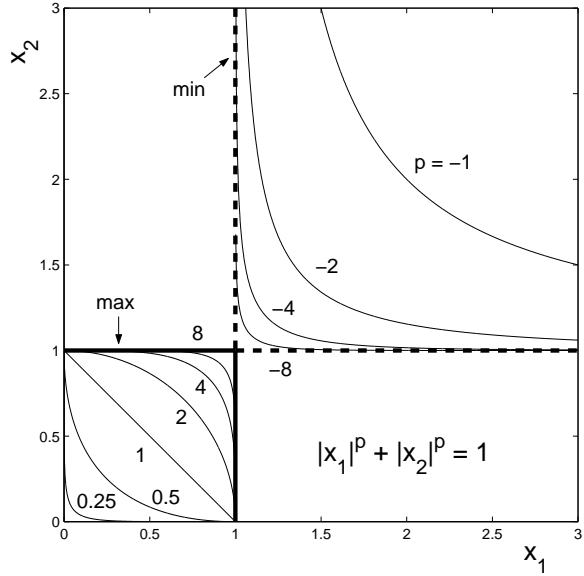} 
    }
\label{fig:isonorms}
\caption{Example of $p$-sizes:
  A vector $\mathbf{A} = (x_1,x_2)$ has many different norms or, more
  generally, $p$-sizes. Conversely, different sets of vectors
  $\mathbf{A}$ have the same $p$-sizes, as shown in this picture. The
  usual quarter-circle Pythagorean norm ($p{=}2$) is complemented by
  various convex and concave ``isonorm'' curves. Of particular
  interest are the $p= +\infty$ (maximum) and $p = -\infty$ (minimum)
  sizes.  }
\end{figure}

While the above result represents a strong endorsement of the Q-space
approach to multiparticle correlations, there is more.
\textit{Generalised $p$-norms} are the extension of Eq.~(\ref{szd}) to
arbitrary $p\geq 1$: Given an $m$-dimensional vector $\mathbf{A} =
(a_1,\dots,a_m)$, the p-norm of $\mathbf{A}$ is defined by
\begin{equation}
  \slabel{sze}
  \pnorm{\mathbf{A}}{p} \equiv 
  \Bigl[\; \sum_{i=1}^m |a_i|^p \,\Bigr]^{1/p} \,.
\end{equation}
Special cases are the $p{=}2$ norm appearing in Eq.~(\ref{szd}) and
the ``max'' norm,
\begin{equation}
\slabel{szf}
\lim_{p\to\infty} \pnorm{A}{p} = \max_i (\,|a_i|\,) \,.
\end{equation}
The condition $p\geq 1$, required for $\pnorm{\mathbf{A}}{p}$ to
satisfy the Minkowski inequality $\pnorm{\mathbf{A} + \mathbf{B}}{p}
\leq \pnorm{\mathbf{A}}{p} + \pnorm{\mathbf{B}}{p}$, can be relaxed to
$p \in \mathbb{R}$ if we do not insist on $\pnorm{A}{p}$ being a norm
and allow it to be merely a ``size measure'' or ``$p$-size''. We then
also have negative-$p$ sizes and in particular
\begin{equation}
\slabel{szg}
\lim_{p\to\; -\infty} \pnorm{A}{p} = \min_i (\,|a_i|\,) \,,
\qquad |a_i| \neq 0\ \forall i \,.
\end{equation}
Fig.~\ref{fig:isonorms} shows as a simple example for a two-component
vector $\mathbf{A} = (x_1,x_2)$ the set of \textit{isonorms}
satisfying $\pnorm{\mathbf{A}}{p} = 1 = \left[\,|x_1|^p +
  |x_2|^p\,\right]^{1/p}$ i.e.\ a set of curves showing which vectors
$\mathbf{A}$ have the same ``$p$-size''. Only positive $x_1$ and $x_2$
are shown as $\pnorm{\mathbf{A}}{p}$ is reflection-symmetric about
$x_1{=}0$ and $x_2{=}0$. The usual circle for constant $2$-norm is
complemented by the straight diagonal line (or, in the full plane,
diamond shape) of $p{=}1$ and various shapes in between. Of particular
interest are the max and min size measures shown as the solid line and
dashed line respectively.
\pp

Based on the above, we can define a ``GHP sum $p$-size'' as follows:
\begin{equation}
\slabel{szp}
S_{q,p}^{{\rm (GHP)}} \equiv 
\Bigl[ \sum_{i < j =1}^q |\psx_i - \psx_j|^p \Bigr]^{1/p}
\,,\quad p\ne 0
\end{equation}
which for $p\geq 1$ is a norm. Eq.~(\ref{szd}) is seen to be the
special case $S_{q,2}^{{\rm (GHP)}} = \sqrt{q} \; \pnorm{\qsx^D}{2}$,
while the GHP max definition is the special case $p \to \infty$,
\begin{equation}
\slabel{szpm}
S_{q,\infty}^{{\rm (GHP)}} = \max_{i < j =1}^q |\psx_i - \psx_j| \,,
\end{equation}
representing the size prescription used in the correlation integral
(\ref{psc}). 
\pp

A given Q-space vector $\mathbf{A}$ will therefore yield a set of size
measures $\left\{ \pnorm{\mathbf{A}}{p}\; | \; p \in
  \mathbb{R},\; p \neq 0 \right\}$.  which includes the ``min'',
``max'' as well as the usual 2-norms. An infinite set of $p$-sizes of
a given $q$-tuple is, however, mostly redundant; for practical
purposes, a subset such as $p \in \{-\infty, 0.5, 2, +\infty\}$ is
probably sufficient.
\pp

The sizes $S_{q,p}$ can be understood in terms of projections in
Q-space as follows. Define the set of \textit{pair plane normal
  vectors} $\hsu_{ij} \equiv (\isu_i - \isu_j)/\sqrt{2}$; these can be
considered to span the respective pair $(i,j)$'s difference space
$\dspace_2^{(i,j)}$. Eq.~(\ref{szp}) then can be written as
\begin{equation}
\slabel{szq}
S_{q,p}^{{\rm (GHP)}} =
\Bigl[ \sum_{i < j =1}^q |\hsu_{ij}\bcdot \qsx^Q|^p \Bigr]^{1/p} \,.
\end{equation}
The set of $\hsu_{ij}$'s are clearly not mutually orthogonal, given
that $q(q-1)/2$ vectors $\hsu_{ij}$ all live in the
$(q-1)$-dimensional difference space $\dspace_q$: indeed, we have
$\hsu_{ij}\bcdot \hsu_{kl} = (\delta_{ik} + \delta_{jl} - \delta_{il} -
\delta_{jk})/2$ and, conversely, Eqs.~(\ref{bsra})--(\ref{bsrc}) show
that the $\hsu_{q\geq2}$ are simple sums of the $\hsu_{ij}$'s.
Fig.~5 shows, for the simple case $q{=}3$ and $d{=}1$, the normal
vectors in the difference space $\dspace_3$ (equivalent to viewing the
$\qspace$ cube of Fig.~\ref{fig:threetuples}(a) down the main
diagonal) as well as the distances from the respective pair planes
$d_{ij} = |\psx_i - \psx_j|/\sqrt{2}$.
\begin{figure}
  \hfil 
  \parbox[c]{55mm}{
    \fbox{\includegraphics[scale=0.30]{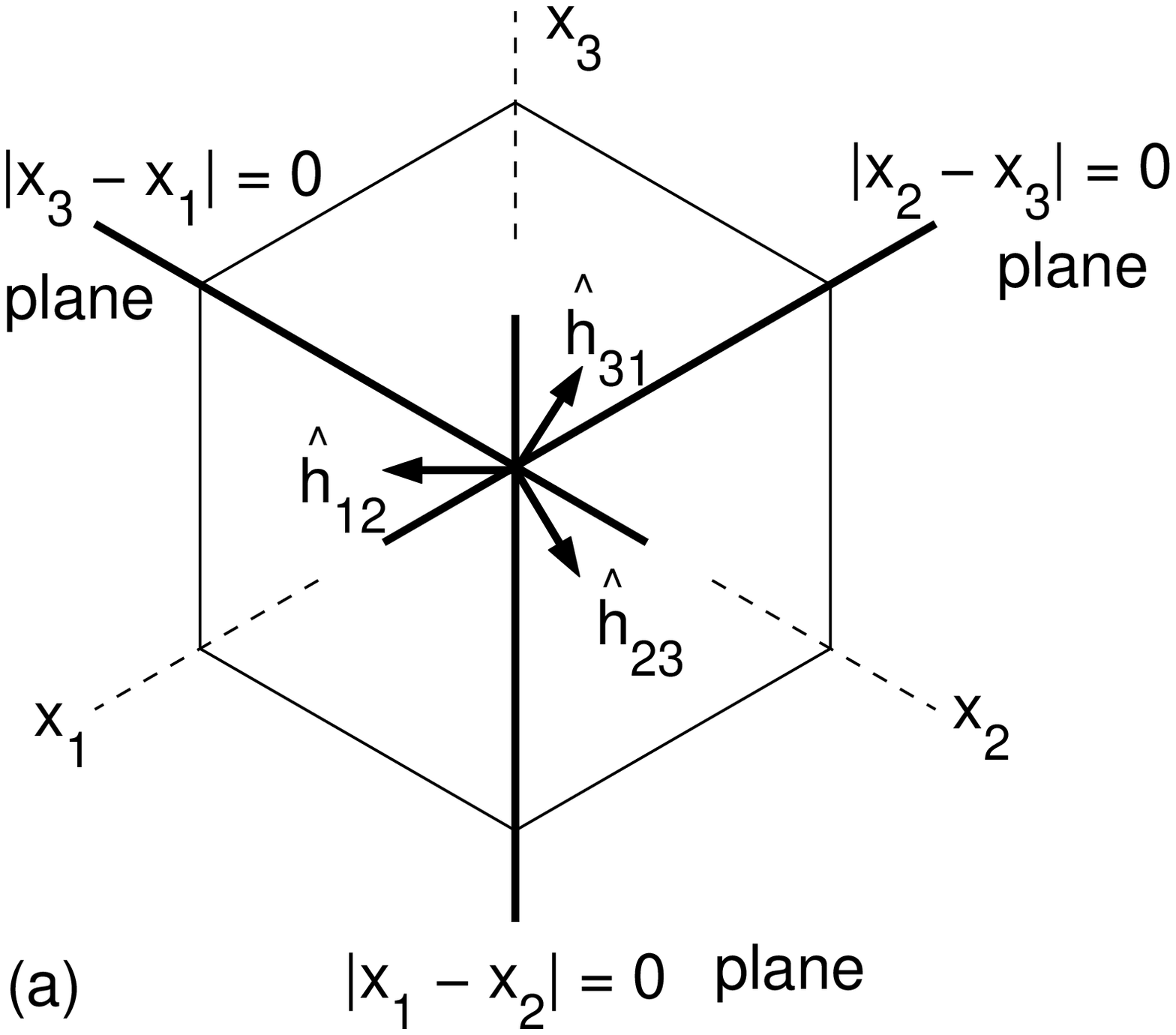} } } 
  \hfil
  \parbox[c]{55mm}{
    \fbox{\includegraphics[scale=0.30]{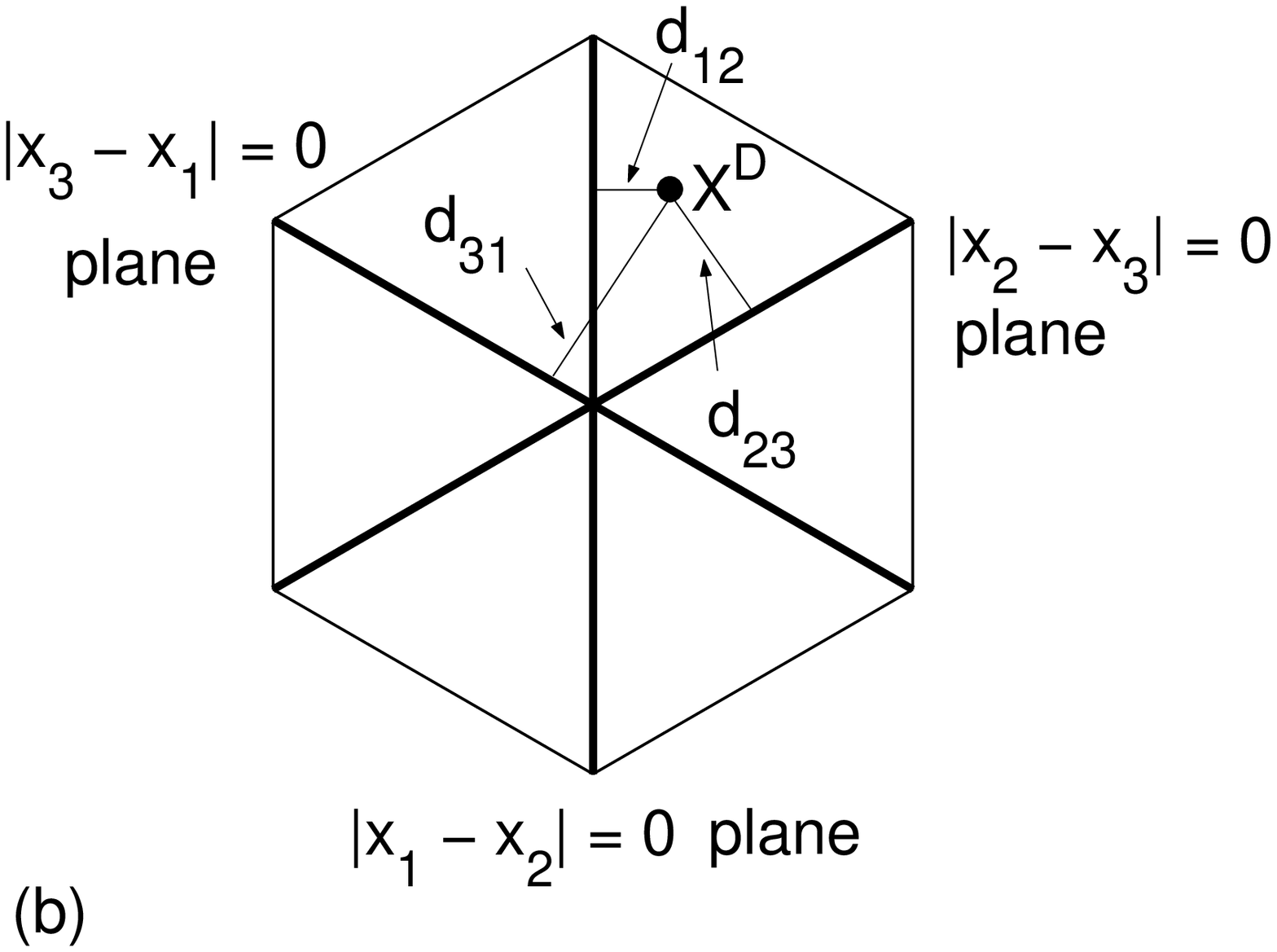} } }
  \hfil
  \\
\label{fig:fpairs}
\caption{Example ($d{=}1$, $q{=}3$) of the nonorthogonal projection (\textit{not}
  decomposition!)  of Q-space vectors in terms onto ``pair plane
  normals'' $\hsu_{ij}$.  The difference space vector $\qsx^D$, shown
  as a point in (b), is projected onto the pair plane normal vectors
  $\hsu_{12}$, $\hsu_{23}$ and $\hsu_{31}$ to yield the distances
  $d_{ij}$ between $\qsx^D$ and the respective pair planes $|\psx_i -
  \psx_j|=0$. The views are once again down the main diagonal.}
\end{figure}

\section{Q-space and other size measures}

The symmetry of Q-space clearly favours the GHP topology over the
corresponding Star and Snake topologies. It is possible, nevertheless,
to accommodate the latter into Q-space.  A simple ad hoc definition of
the Star $p$-size of the $q$-tuple centered on $\psx_i$ would be
\begin{equation}
\slabel{szr}
S_{q,p}^{{\rm (Star)}}(\psx_i) \equiv 
\Bigl[ \sum_{j =1}^q |\psx_i - \psx_j|^p \Bigr]^{1/p}
\,,\quad p\ne 0 \,.
\end{equation}
This is the $p$-generalised size measure corresponding to the
Star max correlation integrals of Eq.~(\ref{psd})--(\ref{pse}). 
Alternatively, we can, for the Star $q$-tuple centered on $i{=}1$,
define a vector $\mathbf{X}^S_{{\rm Star}}(\psx_1) \equiv \sqrt{q}\,
\psx_1 \hsu_1$ and from this find the vector
\begin{equation}
\slabel{szs}
\mathbf{X}^D_{{\rm Star}}(\psx_1)
\equiv \qsx^Q - \mathbf{X}^S_{{\rm Star}}(\psx_1) 
= (\psx_2 - \psx_1)\isu_2 + \cdots + (\psx_q - \psx_1)\isu_q \,;
\end{equation}
then the size measure (\ref{szr}) can be applied directly in terms of
its $\isu_j$ components. Clearly, the pair $(\mathbf{X}^S_{{\rm
    Star}}(\psx_1), \mathbf{X}^D_{{\rm Star}}(\psx_1))$ is not
orthogonal to $\hsu_1$ and hence lives neither in $\sspace_q$ nor
$\dspace_q$; also, this pair will be different for every $q$-tuple
centre $\psx_i$. This is consistent with the fact that sizes of Star
$q$-tuples are, by definition, different for the different centre
particles.
\pp

Star and Snake topologies can be accommodated in the Q-space presented
here, but, due to their obvious asymmetry, do not fit comfortably into
the explicitly symmetric Q-space framework.  Other approaches such as
conditional spaces (for the Star) and spaces based on strict ordering
such as in time series (for the Snake) will probably yield more
natural interpretations for these cases.

\section{Summary}

It is becoming clear that the concept of Q-spaces is yielding insight
into the fundamental structure of correlations of point sets. As
subspaces of the final-state phase space $q = N$, Q-spaces have a
solid theoretical and historical foundation, while offering a
structured set of numbers characterising $q$-tuples, starting with
location and size. Generalising simple notions of norms to an infinite
set of $p$-sizes, we find that the ``max'' and 2-sum sizes are just
special cases within this wider set. Nonorthogonal ``pair planes''
projections are found to be important, in keeping with the obvious
fact that relative distances are also mutually constrained and hence
dependent.
\pp

Most importantly, the usual 2-norm of the difference space vector is
found to be the GHP sum size prescription, which therefore should be
afforded more attention in higher order correlation analyses.
\pp

Extensions such as shape characterisers and higher-order measures of
size are easily conjured up.

\section*{Acknowledgements}

This work was supported in part by the National Research Foundation of
South Africa and by the United States Department of Energy.



\begin{thebibliography}{99}

\bibitem{Egg01a}H.C.\ Eggers,
                in: 30th Int.\ Symp.\ on Multiparticle Dynamics, 
                World Scientific (2001) pp.\ 291--302;
                \textbf{hep-ex/0102005}

\bibitem{Stu87a}A.\ Stuart and J.K.\ Ord, 
               {\it Kendall's Advanced Theory of Statistics}, 
               Vol.1, fifth edition, Oxford University Press, 
               New York (1987).

\bibitem{Lip92a}P.\ Lipa, P.\ Carruthers, H.\ C.\ Eggers and B.\ Buschbeck, 
                Phys.\ Lett.\ B\textbf{285}, 300 (1992);
                H.\ C.\ Eggers, P.\ Lipa, P.\ Carruthers and B.\ Buschbeck, 
                Phys.\ Rev.\ D\textbf{48}, 2040 (1993).

\bibitem{Gra83a}P.\ Grassberger and I.\ Procaccia,
                Phys.\ Rev.\ Lett.\ \textbf{50}, 346 (1983).

\bibitem{Hen83a}Hentschel and Procaccia,
                Physica \textbf{8D}, 435 (1983).

\bibitem{Gra83b}P.\ Grassberger,
                Phys.\ Lett.\ A\textbf{97}, 227 (1983).

\bibitem{Pal84a}G.\ Paladin and A.\ Vulpiani,
                Lett.\ Nuovo Cimento \textbf{41}, 82 (1984).

\bibitem{Paw87a}K.\ Pawelzik and H.\ G.\ Schuster,
                Phys.\ Rev.\ A\textbf{35},481 (1987).

\bibitem{Atm89a}H.\ Atmanspacher, H.\ Scheingraber and G.\ Wiedenmann,
                Phys.\ Rev.\ A\textbf{40}, 3954 (1989).


\bibitem{Foa75a}L.\ Fo\` a, 
                Phys.\ Rep.\ \textbf{22}, 1 (1975).

\end{thebibliography}
\end{document}